\documentclass[10pt]{article}
\usepackage[dvips]{graphicx}
\def\baselinestretch{1.2}
\newcommand\nucl[2]{{}^{#1}{\rm #2}}
\newcommand{\geff}[1]{g_{#1}}

\begin{document}

\setcounter{page}{0}
\vspace{3cm}
\hfill
{\large {\bf PSI\,-\,PR\,-\,00\,-\,14}}\\
\hspace*{\fill}
{\large September 2000}
\vspace{5cm}
\begin{center}
{\Large \bf
 ON THE STRENGTH OF SPIN-ISOSPIN \\ TRANSITIONS
 IN {\boldmath A=28} NUCLEI
}\\
\vspace{2cm}
V.A.\. Kuz'min$^{1}$,
T.V.\ Tetereva$^{2}$ 
and K.\ Junker$^{3}$ \\ 
 {\it
   $1$~Joint Institute for Nuclear Research, Dubna,
   Moscow region, 141980, Russia} \\ 
 {\it
 $2$~Skobeltsyn Institute of Nuclear Physics,
   Lomonosov Moscow State University, Moscow, Russia} \\ 
 {\it
 $3$~Paul Scherrer Institut, CH-5232 Villigen-PSI, Switzerland} \\ 
\end{center}
\vspace{6cm}
\hspace*{9cm}
{\large 
PAUL SCHERRER INSTITUT} \\
\hspace*{9cm} {\large CH - 5232 Villigen PSI} \\
\hspace*{9cm} {\large Telephon 0041 56 310 2111} \\
\hspace*{9cm} {\large Telefax 0041 56 2199}\\*[1cm]
{\small
Paper presented at the\\
International Conference on Nuclear Structure and Related Topics,\\
June 2000, Dubna, Russia}
\newpage
\def\baselinestretch{1.2}
\begin{center}
{\large \bf
 ON THE STRENGTH OF SPIN-ISOSPIN \\ TRANSITIONS
 IN {\boldmath A=28} NUCLEI
}

\medskip
V.A.\. Kuz'min$^{1}$,
T.V.\ Tetereva$^{2}$ 
and K.\ Junker$^{3}$ \\ 
 {\it
   1~Joint Institute for Nuclear Research, Dubna,
   Moscow region, 141980, Russia} \\ 
 {\it
 2~Skobeltsyn Institute of Nuclear Physics,
   Lomonosov Moscow State University, Moscow, Russia} \\ 
 {\it
 3~Paul Scherrer Institut, CH-5232 Villigen-PSI, Switzerland} \\ 
\end{center}
\begin{abstract}
 The relations between the strengths of spin-isospin transition
 operators extracted from direct nuclear reactions,
 magnetic scattering of electrons and processes of semi-lepton
 weak interaction are discussed.
\end{abstract}

\section{Introduction}

 The studies of the spin-isospin excitations in atomic nuclei
 have a long history.
 Detailed discussions of it are given by the authors of recent
 reviews \cite{Osterfeld1992,AS1998}.
 We touch here only a few points important for our
 purposes.
 The first manifestation of spin-isospin transitions was detected
 in beta-decay as Gamow-Teller transitions
 ($\Delta J^{\pi} = 1^{+}$) for which
 $$
  \log f t_{1/2} = {6135 \over (\geff{A} / \geff{V})^{2} B^{\pm}(GT)}.
 $$
 Here the strength of the Gamow-Teller transition is introduced as
 \begin{equation}
 \label{eq:bgt}
   B^{\pm}_{f}(GT) = { 1 \over 2 J_{f} + 1 }
   \biggl \vert
   \langle J_{f} \biggl \Vert \sum_{k=1}^{A} \sigma_{k} \, t^{\pm}_{k}
   \biggr \Vert J_{i} \rangle \biggr \vert^{2} \, .
 \end{equation}
 In this paper we will only discuss the transitions
 $0^{+} \rightarrow 1^{+}$.
 The discovery of isobar analog states in $(p,n)$ reactions was followed by
 the prediction of the a new nuclear collective excitation -- the 
 giant GT resonance -- as the reason of lack of $\sigma t^{-}$ strength
 observed in $\beta$-decay studies.
 At the beginning of 1980-s giant GT resonances were experimentally
 discovered and studied in $(p,n)$ and other nuclear charge-exchange 
 reactions (CEX) at intermediate energies.
 Using some additional assumptions, it was shown in \cite{Goodman1980} 
 that the \hbox{$0^{\circ}\, (p,n)$} cross sections are proportional to $B(GT)$.
 The comparison of the $B(GT)$ values extracted from the cross sections of
 CEX reactions with those obtained from $\beta$-decay reveals that
 some differences exist between them \cite{diff}.
 The origin of these differences has been explained by the fact that
 transitions with small $B(GT)$ are observed even in fast beta decay.
 For small $B(GT)$, however, other spin-multipoles contribute strongly to the
 CEX cross sections leading to a considerable deviation from the
 proportionality between the $0^{\circ}$ cross sections and the $B(GT)$ values.
 Therefore large errors may appear in $B(GT)$'s
 obtained from the CEX cross sections \cite{AAL1994}.

 Recent experiments on exclusive muon capture in $sd$-shell
 nuclei \cite{Gorringe1999} give the possibility to compare the
 characteristics of strong GT transitions measured in weak
 interaction processes to those obtained from the CEX reactions.
 The energy released in nuclear muon capture is determined by the muon
 mass. Limitations on the transition energy, that exist in
 beta decay, are absent in muon capture.
 During muon capture the nucleus acquires a non-zero linear momentum.
 Therefore the kinematics in muon capture differs from that
 in beta decay and zero-angle CEX reactions. For this reason the matrix  
 elements for $0^{+} \rightarrow 1^{+}$ transitions obtained in muon 
 capture can not be compared directly to
 the $B(GT)$'s extracted from $(p,n)$ reactions.
 One is therefore forced to ask a different question, namely, as to what
 extend the wave functions of the isovector states will simultaneously describe
 the experimental $B(GT)$'s and the rates of ordinary muon capture 
 ($\Lambda_{f}$).

\section{The nuclear muon capture rate}

 We base our calculations of exclusive muon capture rates on the
 approach described in \cite{BE1967}.
 In case that the matrix elements of the operator
 $ \displaystyle \sum_{k=1}^{A}
 j_{0} (E_{\nu} r_{k}) \, \sigma_{k} \, t^{+}_{k} $
 dominates, the rates for the partial transitions
 $0^{+}_{\rm g.s.} \rightarrow 1^{+}_{f}$ are give by (only the final result
 is shown here)
 
 \begin{equation}
 \label{eq:rate}
  \matrix{
  \Lambda_{f} \approx & \displaystyle {2 \over 3} \, V \,
   g^{2}_{A} \, [101]^{2} \, \biggl \lbrace
   1 + { 2 \over 3 } \eta
   + { 8 \over 3 } {{g_{V} + g_{M}} \over g_{A}} \eta
    - {2 \over 3}{g_{P} \over g_{A} } \eta
    + {1 \over 3} \bigl ( {g_{P} \over g_{A} } \eta \bigr )^2
   \hfill \cr
 \noalign{\medskip}
  & \displaystyle
  + \, \sqrt{ 8 \over 9 }
    \biggl \lbrack
    2 \bigl ( 1 + { {g_{V} + g_{M}} \over g_{A} }
    - { g_{P} \over g_{A} } \bigr ) \eta
    + \bigl( { g_{P} \over g_{A} } \eta \bigr )^2
    \biggr \rbrack \, { {[121]} \over {[101]} }
  \hfill \cr
  \noalign{\medskip}
  & \displaystyle
  + \, 2 \, \bigl ( 1 - { g_{P} \over g_{A} } \bigr ) \, \eta
    \, { [111p] \over M \, [101] }
  - \sqrt{ 8 \over 9 } \,  { g_{V} \over g_{A} }
     \, { [011p] \over M \, [101] }
     \biggr \rbrace \, , \hfill \cr
 }
\end{equation}
 where
 $ \displaystyle \eta = { E_{\nu} \over 2 \, M_{p} } $
 and the nuclear matrix elements are defined by
 $$
 \matrix{
 [101] & = \displaystyle
    \sqrt{ 1 \over 4 \pi }
    \langle 1^{+}_{f} \parallel \sum_{k = 1}^{A}
       \varphi_{\mu}(r_{k}) \,
    j_{0}(E_{\nu} r_{k}) \, Y_{0}(\hat{r}_{k}) \, \sigma_{k} \,
    t^{+}_{k}
    \parallel 0^{+}_{\rm g.s.} \rangle  \, , \hfill \cr
 [121] & = \displaystyle
    \sqrt{ 1 \over 4 \pi }
    \langle 1^{+}_{f} \parallel \sum_{k = 1}^{A}
       \varphi_{\mu}(r_{k}) \,
      j_{2}(E_{\nu} r_{k}) \,
      \bigr [ Y_{2}(\hat{r}_{k}) \otimes \sigma_{k} \bigr ]_{1} \,
       t^{+}_{k}
    \parallel 0^{+}_{\rm g.s.} \rangle \, , \hfill \cr
 [111p] & = \displaystyle
    \sqrt{ 1 \over 4 \pi }
    \langle 1^{+}_{f} \parallel \sum_{k = 1}^{A}
       \varphi_{\mu}(r_{k}) \,
     j_{1}(E_{\nu} r_{k}) \, \bigr [ Y_{1}(\hat{r}_{k})
     \otimes  \nabla_{k} \bigr ]_{1} \, t^{+}_{k}
    \parallel 0^{+}_{\rm g.s.} \rangle \, , \hfill \cr
 [011p] & = \displaystyle
    \sqrt{ 1 \over 12 \pi } %
    \langle 1^{+}_{f} \parallel \sum_{k = 1}^{A}
       \varphi_{\mu}(r_{k}) \,
      j_{1}(E_{\nu} r_{k}) \,
       Y_{1} (\hat{r}_{k}) \, \bigr (
       \vec{\nabla}_{k}, \vec{\sigma}_{k} \bigr ) \,
       t^{+}_{k} \parallel 0^{+}_{\rm g.s.} \rangle \, . \cr
 }
 $$
 Here $\varphi_{\mu}(r)$ is muon radial wave function.
 We approximate $\varphi_{\mu}$, as is done usually for 
 light and medium nuclei, by the average value calculated
 in \cite{FW1962}.

 \section{Comparison between calculations and \\experimental data}

\begin{figure}
 \includegraphics{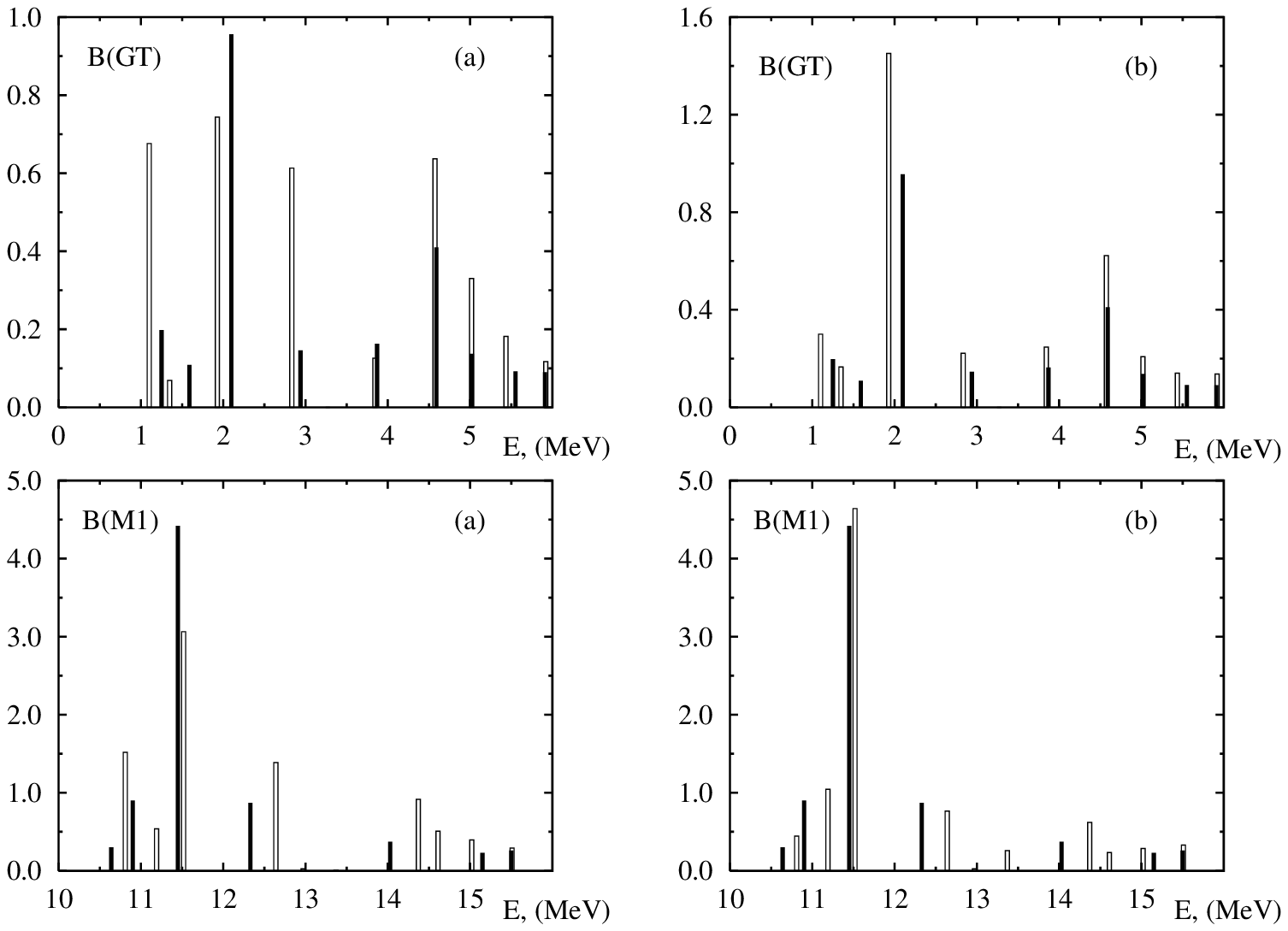}
%
\caption{%
 Strength functions of GT and $M1$ transitions in
 $\nucl{28}{Si}$.
 The experimental data are shown by closed bars;
 the results of calculations -- as open bars.
 The calculations (a) have been carried out with eigenfunctions
 of the Hamiltonian \protect\cite{Wildenthal},
 (b) -- with the transformed wave functions.}
\label{fig:28si}
\end{figure}

 The calculations have been carried out on the basis of a many-particle
 shell model using the  Hamiltonian of B.H.~Wildenthal \cite{Wildenthal}
 and an unrestricted $sd$-shell space.
 The computer code OXBASH \cite{OXBASH} has been used in the calculations. 
 Theoretical and experimental GT- and $M1$-strength functions are
 presented in \hbox{Fig.\ \ref{fig:28si}}. The 
 theoretical results obtained with the eigenfunctions of Wildenthal's
 Hamiltonian are marked by (a).
 The $B(GT)$ for the $1^{+}$ states with excitation energies below
 \hbox{$6$ MeV} are shown in the upper left part of
 \hbox{Fig.\ \ref{fig:28si}}.
 The energies are measured from the ground state of $\nucl{28}{P}$.
 The experimental GT strength function was obtained from
 the cross sections for the
 \hbox{$\nucl{28}{Si}(p,n)\nucl{28}{P}$} reaction \cite{si28pn}.
 All states with excitation energy below \hbox{$5$ MeV} are
 shown in \hbox{Fig.\ \ref{fig:28si}}.
 Only a small fraction of the whole experimental GT strength
 goes to the states with higher excitation energies \cite{si28pn}.
 Additionally, the exact spins and parities of high-lying states
 have not been determined experimentally.
 Because of these two reasons we neglect in the following considerations 
 the high-lying states which are not shown in \hbox{Fig.\ \ref{fig:28si}}. 
 In the energy region up to \hbox{$12.6$ MeV} one observes an 
 experimental $B(GT)$ strength of $2.595$; 
 the $B(GT)$'s summed over the states shown in \hbox{Fig.\ \ref{fig:28si}}
 amounts to $2.301$.
 The theoretical $B(GT)$'s summed over the first 10 eigestates with
 $J^{\pi},T = 1^{+},1$ (shown in \hbox{Fig.\ \ref{fig:28si}})
 give $3.492$. Therefore a rather standard value of GT quenching
 $ \displaystyle {\Sigma B_{\rm exper.}(GT) \over
   \Sigma B_{\rm theor.}(GT)} =  0.66 $
 is obtained from this comparison.

 Figure \ref{fig:28si} shows also the theoretical and experimental
 $M1$ strength functions.
 It is known \cite{EndtBooten} that the shell model with the Hamiltonian
 \cite{Wildenthal} reproduces well the energies of isovector
 $1^{+}$ states in $\nucl{28}{Si}$.
 But the theoretical dependence of the $B(M1)$ values on excitation energy
 differs considerably from the experimental one, obtained in
 \cite{si28ee}.
 The calculations were carried out with the ``free'' value of $g_{s}$
 and the summed theoretical $B(M1)$ is larger than the experimental
 one : $ \displaystyle
 {\Sigma \, B_{\rm exper.}(M1) \over \Sigma \, B_{\rm theor.}(M1) } =
 { 7.360 \over 8.623 } = 0.85 $.
 However, as can be seen in \hbox{Fig.\ \ref{fig:28si}},
 even in that case the experimental $B(M1)$ exceeds
 considerably the theoretical value for the strongest transition
 which goes to the isovector $1^{+}$ state with energy \hbox{$11.445$ MeV}.

 Therefore we can conclude that the shell model with the Hamiltonian
 \cite{Wildenthal} describes qualitatively the main features of GT
 and $M1$ strength functions in the sense that small theoretical
 $B(GT)$'s and $B(M1)$'s correspond to small experimental
 values.
 However, the theoretical distributions of the transition strength over
 the states which absorb the largest part of the total strength
 differ considerably from the experimental strength functions.

 According to (\ref{eq:rate}), the nuclear matrix element $[101]$,
 having the $\sigma t^{+}$ operator as the spin-angular part, contributes
 mainly to the rate in fast allowed muon capture.
 Therefore, the differences between theoretical and experimental
 values of $B(GT)$ and $B(M1)$ led to the discrepancies between
 theoretical and experimental values of $\Lambda_{f}$'.
 The $\Lambda_{f}$'s calculated with the eigenfunctions of the Wildenthal
 Hamiltonian are shown in \hbox{Table \ref{tab:transitions}} in
 column (a).
 Also, the values of $B(M1)$ and $B(GT)$ for the members of
 the same isotopic triplets are presented in the Table together
 with the corresponding experimental numbers.
 The only conclusion, which one can make comparing experimental data
 to results of calculation (a), is that the difficulties in description
 of GT and $M1$ transitions have there counterpart in the description
 of muon capture rates.

 In this situation it might be reasonable to use the available
 experimental information concerning GT and $M1$ strength functions
 in the calculations of muon capture rates.
 For that purpose the orthogonal transformation, acting in the subspace
 spanned by the wave functions of the isovector $1^{+}$ states, was
 suggested in \cite{KT1999}.
 According to this paper
 the parameters of the transformation should be chosen such that
 the strength functions of GT and $M1$ transitions calculated with
 transformed wave functions coincide in shape (up to a constant factor)
 with the experimental GT and $M1$ strength functions.
 Therefore, the transformation parameters do not dependent on any
 relation between the experimental and theoretical values of the
 summed strengths of GT and $M1$ transitions and are determined
 by the shapes of experimental GT and $M1$ strength functions only.
 The orthogonality of the transformation will support the mutual
 orthogonality and normalization of the obtained wave functions.
 For the same reason, the theoretical total GT and $M1$ transition
 strength will be conserved.

 The isotopic invariance of strong interactions assures that
 the transformation of the isovector $1^{+}$ states carried out
 in $\nucl{28}{P}$ will induce the transformations of the
 $1^{+},1$ states in $\nucl{28}{Si}$ and $\nucl{28}{Al}$.
 In addition, the transformation matrix will not depend on the
 value of the third component of the total isospin.
 Therefore exactly the same transformation will apply for the corresponding 
 subspaces of the wave functions 
 for $\nucl{28}{Si}$ and $\nucl{28}{Al}$.

 The next section describes how this transformation can be
 constructed.

\section{Transformation of wave functions}

 The transformation of wave functions of excited states
$$
 \phi_{k} \rightarrow
 \psi_{k} = U_{k,k^{\prime}} \, \phi_{k^{\prime}}
 \qquad ( k = 1, 2, \ldots, N )
$$
 causes a transformation of the transition matrix elements
$$
 \langle \phi_{k} \vert {\cal O} \vert \Phi \rangle \rightarrow
 \langle \psi_{k} \vert {\cal O} \vert \Phi \rangle =
   U^{\ast}_{k,k^{\prime}} \, \langle \phi_{k^{\prime}} \vert
  {\cal O} \vert \Phi \rangle =
  \langle \phi_{k^{\prime}} \vert {\cal O} \vert \Phi \rangle
  \, U^{\dag}_{k^{\prime},k} .
$$
 Considering the transformation within the subspace of the multiparticle
 wavefunction as a transformation in a vector space with the transition
 amplitudes as basis vectors will simplify considerably the determination
 of the transformation matrix.

 An orthogonal $N \times N$ matrix is determined through
 $ N(N-1) / 2 \, $ free parameters.
 To reduce the number of required parameters one should use
 matrices of less general structure.
 The simplest orthogonal transformation of a vector is its reflection
 on a plane \cite{Cartan,Householder}
$$
  v = v_{\parallel} + v_{\perp} \rightarrow
   v^{\prime} = v_{\parallel} - v_{\perp} \ ,
$$
 Here the vector $v_{\parallel}$ is parallel to the plane and
 $v_{\perp}$ is perpendicular to the plane.
 If the plane is determined by the equation
      $ (b,x) = b_{k} x_{k} = 0, $
 where $b$ is a non-zero vector, $ (b,b) > 0, $
 then according to \cite{Cartan,Householder} the transformation
 is given by
\begin{equation}
\label{eq:refl}
   v_{k} \rightarrow
   v^{\prime} = R(b) \, v \quad  {\rm with}  \quad
   R_{k,l}(b) = \delta_{k,l} - 2 \, { b_{k} b_{l} \over (b,b) } \ .
\end{equation}
 If $\vert u \vert = \vert w \vert$ one can convert
 $ u $ into $ w $ and vice versa by the
 transformation (\ref{eq:refl}) with $ b = u - w $.

 From the calculations within the shell model we know the vector,
 built up from the theoretical GT amplitudes. A
 second vector is assembled from the experimental amplitudes.
 After proper normalization the matrix (\ref{eq:refl}) can
 be constructed.
 However from the experiment we can get only the absolute values
 of the transition amplitudes. Therefore we are forced to consider all
 possible distributions of signs within the ``experimental vector''.
 Each distribution has its own reflection matrix, which
 will transform the wave functions in such a way that the new
 theoretical GT strength function will coincide in shape with the
 experimental one.
 In order to select the best transformation we consider the
 $M1$ strength function in $\nucl{28}{Si}$.
 Due to isotopic invariance the 
 $1^{+}$ states in $\nucl{28}{Si}$ and $\nucl{28}{P}$ are transformed by
 the same matrix. 
 Therefore, the theoretical $M1$ strength function will be changed
 too.
 The magnetic dipole transition operator differs from the GT transition
 operator and
 the vectors built up of GT and $M1$ amplitudes will be
 linearly independent.
 Therefore, the transformed theoretical $M1$ strength function
 will have a different shape from the experimental one.
 We use that transformation which leads to the smallest deviation
 between the theoretical and experimental $M1$ shapes.

\section{Calculations with transformed wave functions}

 The GT and $M1$ strength functions calculated with the transformed
 wave functions are presented on the right hand side of
 \hbox{Fig.\ \ref{fig:28si}}, marked by (b).
 The subspace, in which the transformation acts,
 includes all the states shown in \hbox{Fig.\ \ref{fig:28si}}.
 The transformation causes a significant redistribution
 of transition strengths over the excitation energies.
 As a result, the shape of GT strength function
 is exactly restored and the shape of the $M1$ strength function
 is approximately reproduced.
 The muon capture rates calculated with the transformed wave
 functions are given in column (b) of \hbox{Table \ref{tab:transitions}}.
 The new theoretical rates are very closed to the experimental ones.
 The errors in $\Lambda_{f}$ are estimations of the uncertainties
 in the calculated rates induced by errors in the experimental values of 
 $B(GT)$'s and $B(M1)$'s, which were used in the construction of the 
 transformation.
 It should be pointed out again that the experimental values of $B(GT)$
 and $B(M1)$ themselves have not been used in transformation matrix,
 only the shapes of the experimental GT and $M1$ strength functions were
 important for the transformation.
 Also, no effective charges were introduced in the calculations.
 It is reasonable to compare $B(GT)$ and $B(M1)$ calculated with the
 transformed wave functions to the experimental values.
 The result is given in column (b) of
 \hbox{Table \ref{tab:transitions}}.

 The calculations with the transformed wave functions, carried out
 for the strongest transitions, produce surprising results:
 the theoretical OMC rates are very close to the experimental values;
 the theoretical $B(M1)$'s are close to the experimental ones;
 but the theoretical $B(GT)$'s are $1.5$ times larger than those
 extracted from the cross sections of reaction
 $\nucl{28}{Si}(p,n)\nucl{28}{P}$.
 Because of the similarity of the spin-isospin parts of the
 operators describing CEX reactions, magnetic scattering of electrons
 and muon capture, this disagreement is unexpected.

\section{Conclusions}

 We have constructed a set of wave functions of the excited isovector
 $1^{+}$ states in $A=28$ nuclei starting from the wave functions
 calculated within a many-particle shell model using the Hamiltonian
 \cite{Wildenthal} and
 introducing phenomenological corrections by means of an
 orthogonal transformation in a subspace of shell model wave functions.
 Then, several characteristics of the spin-isospin transitions
 were calculated with the new wave functions.
 The calculations were carried out without introducing any effective
 charges.
 The theoretical results being compared to experimental data
 show that for the strongest isovector $0^{+} \rightarrow 1^{+}$
 transitions:
 i) the theoretical OMC rates are very close to the experimental values;
 ii) the theoretical $B(M1)$'s are close to the experimental ones;
 iii) the theoretical $B(GT)$'s are $1.5$ times larger than those
 extracted from the cross sections of the reaction
 $\nucl{28}{Si}(p,n)\nucl{28}{P}$.
 This disagreement is unexpected mainly due the to similarity of
 spin-isospin parts of the operators describing CEX reactions,
 magnetic scattering of electrons and muon capture.

 We have shown that experimental data on partial muon
 capture rates can be used to obtaine important spectroscopic
 information, because fast spin-flip transitions were observed
 and the rates of weak interaction processes have been measured for
 such fast transitions.

 In contrast to the general accepted opinion the relation
 between cross sections of CEX reactions
 and $B(GT)$ could be quite complicated even for the strong
 GT transitions. It seems to be necessary to
 investigate how spin-quadrupole transitions and two-step
 processes could contribute to cross sections of CEX reactions
 even for strong GT transitions.

\newpage

\begin{table}
\caption{%
 Properties of spin-isospin transitions in $A=28$ nuclei.
 References to experimental data and details of the calculations
 are given in the text.}
\label{tab:transitions}
$$
\begin{tabular}[t]{| c | c | c | c |}
\hline
 $ E_{f} $ & experiment &
 \multicolumn{2}{| c |}{ calculations } \\
 \cline{3-4}
  MeV    & {}  & \quad (a) \quad {} & (b) \\
\hline
\noalign{\smallskip}
 \multicolumn{4}{ c }{
 $\Lambda_{f}$ (in $10^{3} \, {\rm s}^{-1}$) \ %
  for $\nucl{28}{Si}(0^{+}_{\rm g.s.}) \, (\mu,\nu) %
  \,\nucl{28}{Al}(1^{+}_{f}) $ } \\
\noalign{\smallskip}
\hline
 $ 1.62 $ & $ 12.9 \pm 2.1 $ & $  3.1 $ & $  7.6 \pm 0.2 $ \\
 $ 2.20 $ & $ 62.8 \pm 7.4 $ & $ 34.1 $ & $ 63.6 \pm 2.4 $ \\
 $ 3.11 $ & $ 14.7 \pm 2.6 $ & $ 26.1 $ & $ 11.2 \pm 0.5 $ \\
\hline
\noalign{\smallskip}
 \multicolumn{4}{ c }{
  $B_{f}(M1)$ (in $\mu_{N}$) %
  for $\nucl{28}{Si}(0^{+}_{\rm g.s.}) \, (e,e^{\prime}) %
  \,\nucl{28}{Si}(1^{+}_{f}) $ } \\
\noalign{\smallskip}
\hline
 $ 10.90 $ & $ 0.90 \pm 0.02 $ & $ 0.538 $ & $ 1.044 $ \\
 $ 11.45 $ & $ 4.42 \pm 0.20 $ & $ 3.064 $ & $ 4.461 $ \\
 $ 12.33 $ & $ 0.87 \pm 0.06 $ & $ 1.387 $ & $ 0.764 $ \\
\hline
\noalign{\smallskip}
 \multicolumn{4}{ c }{
  $B^{-}_{f}(GT)$ %
  for $\nucl{28}{Si}(0^{+}_{\rm g.s.}) \, (p,n) %
  \, \nucl{28}{P}(1^{+}_{f}) $ } \\
\noalign{\smallskip}
\hline
 $ 1.59 $ & $ 0.109 \pm 0.002 $ & $ 0.069 $ & $ 0.165 $ \\
 $ 2.10 $ & $ 0.956 \pm 0.005 $ & $ 0.774 $ & $ 1.451 $ \\
 $ 2.94 $ & $ 0.146 \pm 0.003 $ & $ 0.613 $ & $ 0.222 $ \\
\hline
\end{tabular}
$$
\end{table}

\clearpage

\end{document}